\documentclass{jaa}
\usepackage{txfonts}
\usepackage{graphicx}
\newcommand{\newblock}{}
\usepackage[authoryear]{natbib}
\begin{document}
%                 So begins the second great struggle...                %

\title[]
{Optical Monitoring of OT 546 in 2009 }
\author[Guo et al. ]
       {Difu Guo%
       \thanks{e-mail:difu@sdu.edu.cn},     Shaoming Hu \& Xu Chen\\
     School of Space Science and Physics, Shandong University, Weihai,\\
        180 Cultural West Road, Weihai, Shandong 264209, China.
   }\maketitle
\label{firstpage}

\begin{abstract}
We reported the monitoring results of OT 546 in V, R and I bands,
observed on 22 nights from February 16 to July 1 in 2009 at Weihai
Observatory of Shandong University. During our monitoring, its
variability amplitude was small and a possible microvariability was detected
on one night using both C and F tests.

\end{abstract}


\section{Introduction}

OT 546 (ZW I 187) was classified as a BL Lac object by
\cite{Angel80}. An optical variation of 2.1 mag was found
in $B$ band \citep{Hall72}, but its variability
amplitude was small recently. The average brightness in V and B bands
were near 16 mag \citep{Kinman76} and 16.7 mag \citep{pica88} respectively.
Rapid short term variability was reported by \cite{pica88}.

\section{Observations and results}
The optical observations were carried out at Weihai Observatory
 of Shandong University, using 1.0-m telescope equipped with a  back
 illuminated PIXIS 2048B CCD camera at the cassegrain focus. The
 field of view was about 12$'$ $\times$ 12$'$. Standard Johnson Cousins
 filters were used in our observations.
The data reduction was carried out following the standard procedure
of the IRAF packages. All images were processed
by bias and flat-field correction. Then, aperture
photometry was used to obtain the instrumental magnitudes for the
 target and the comparison stars. Comparison stars B, H and L taken from
 \cite{Fiorucci96} were used, and the magnitude of the source was
 derived by differential photometry.

  Light curves of the source in V, R and I bands were shown in Figure 1.
  During our observations, the variability
  amplitude was small, similar to recent
  observations \citep{Fiorucci96, Katajainen00, xie04}.
  The brightness variation in V band was between 15.72 (JD 2454978.142)
  and 16.05 (JD 2454907.304) mag with an average brightness of
  15.92 mag. The observed largest intra-night variability amplitudes in V,
  R and I bands were $\Delta$$V$ $\sim$ 0.259 mag, $\Delta$$R$ $\sim$ 0.310 mag
  and $\Delta$$I$ $\sim$ 0.237 mag respectively.

\begin{figure}
\centering
\includegraphics[width=12cm, angle=0]{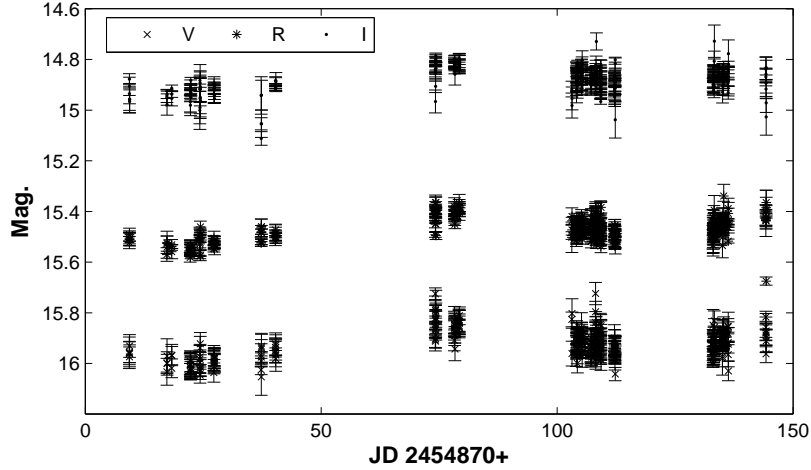}
\caption{Light curves of OT 546 in V, R and I bands. }
\label{fig:pjet}
\end{figure}

  In order to detect the microvariability, both C and F tests have been
  performed to light curves on these 22 nights. Microvariability was detected only on 1 out of 22 nights,
  and the C values \citep{Jang97,Diego10} in V and R bands were 2.767
  and 3.271 respectively.
  At the same time, results of the F-test \citep{Diego10,Gaur12}
  in V and R bands also exceeded the critical F value (1\% significance level)
  on 2009 April 22. So the microvariability is reliable.

\noindent{{\textbf{Acknowledgements:}}} \\
\noindent{This work was partially supported by the NSFC (Nos. 11143012,
 11203016, 10778619 and 10778701), and by the NSF of Shandong Province (No. ZR2012AQ008).

\label{lastpage}

\end{document}